\documentstyle[aps]{revtex}
\newcommand{\Tr}{\mathop{\rm Tr}\nolimits}

\newcommand{\ints}{\int\limits}

\title{\bf  Effective Action for an $SU(2)$ Gauge Model with a Vortex}
\author{V.~Ch.~Zhukovsky\footnotemark[1]\\ {\sl Institut f\"ur Theoretische Physik,
Universit\"at T\"ubingen\\ D-72076 T\"ubingen, Germany}}

\newcommand{\be}{\begin{equation}}
\newcommand{\ee}{\end{equation}}
\begin{document}
\maketitle \large
\begin{abstract}

Effective action of an $SU(2)$ gauge model with a vortex in 4-dimensional
space time is
calculated in the 1-loop approximation. The minimum of the effective
potential is found.
\end{abstract}

\renewcommand{\thefootnote}{\fnsymbol{footnote}}

\footnotetext[1]{\noindent On leave of absence from the
Faculty of Physics, Department of Theoretical Physics,
Moscow State University, 119899, Moscow, Russia.}

\vspace{0.3cm}
\large

\renewcommand{\thefootnote}{\arabic{footnote}}
\setcounter{footnote}{0}
\setcounter{page}{1}
\section{Introduction}
The phenomenon of confinement
in non-Abelian gauge theories has satisfactory
explanation in a number of
models (for a review, see, e.g., \cite{sim}). According to one of them,
the color electric flux of
the quark anti-quark pair is squeezed into a flux tube and this leads
to a linear rise of the quark interaction potential energy. The so
called projection techniques, developed since the early papers of 't
Hooft and Mandelstam \cite{thomand}, provided convenient tools, such
as maximal Abelian gauge, for
further studies of this mechanism on the lattice, which brought
numerical evidence that color electric flux tubes are formed due to the
dual Meissner effect generated by the
condensation of color monopoles \cite{Sch} (see, also  \cite{baker},
\cite {polikar}
and references therein) -- the phenomenon dual to the well known
formation of strings with a magnetic flux in them in the theory of
superconductivity (Abrikosov-Nielsen-Olesen strings
\cite{abrik}). Another explanation of confinement is based on the
method of maximal center gauges \cite{deb}, \cite{langf}, which
displayed the center dominance and lead to the center vortex picture of
confinement. In this picture, the
existence of $Z(N_c)$ vortices, whose distribution in space time fluctuates
sufficiently randomly,  provides for
the so called area decay law for the Wilson loop expectation value,
which implies a linear static quark potential (for  latest references
see, e.g. \cite{lang} ).

There are many questions
left concerning the dynamics of vortices, as well as their
origin. A field theoretical description of them has been started as
early, as in 1978 \cite{gth},  and in the well known ``spaghetti
vacuum'' picture \cite{niels}, which has been recently developed in the
framework of the theory of 4D surfaces and strings in a number of
papers \cite{corn}, \cite{baker}, \cite{antonov}. Recently, the
continuum analogue of the maximum center gauge was constructed and
discussed in \cite{reineng}. At the same time, study of simple models
that allow for exact solutions may shed  light on the complicated
general problem of vortex dynamics.

In the present letter, we consider the one-loop contribution of gauge field
fluctuations about  a pure gauge configuration with a gauge field
vortex in a 4-dimensional space of nontrivial topology, $S^1 \times
R^3$, i.e., with a cylinder. This configuration resembles the simplest
imitation of the Aharonov-Bohm effect with a string and a non-zero
magnetic flux in it.  We demonstrate that our result, with an evident
reinterpretation, confirms the conclusions of
the recent  work of D. Diakonov \cite{diak},  who investigated
potential energy of vortices in the 4- and 3-dimensional gauge
theories.

\section{The effective action}

We consider the generating functional of the gluodynamic model for the
gauge group $SU(2)$ in the 4-dimensional space time $S^1\times R^3$
with a cylinder
\begin{eqnarray}
Z [\bar A, j] = \int d a_\mu^a d \chi
d \bar \chi \exp \left[ -S_4 \right]\int d a_\mu^a \exp \left[ -S_2
\right],
\label{func}
\end{eqnarray}
where the action $S$ of the gauge
field is split into two parts,
\begin{eqnarray}
S = S_4 +S_2=\int d^4x
 (L_4+j^{a \mu} a^a_\mu)+ \int d^2x (L_2+j^{a \mu}
a^a_\mu),
\label{act}
\end{eqnarray} $S_4$ for the 4-dimensional space
time $S^1\times R^3$ and $S_2$  for the 2-dimensional cylinder
$S^1\times R^1$, whose presence, in the same way as a deleted point in
the 2-dimensional plane,  attributes a nontrivial topology to this
space configuration.  The Lagrangian of the gauge field $A_\mu$ in the
external (background) field has the familiar form
\begin{eqnarray}
L_4=\frac{1}{4} (F^a_{\mu\nu})^2 +\frac{1}{2\xi} (\bar D^{ab}_\mu
a^{b\mu})^2 +\bar \chi_a (\bar D^2)_{ab} \chi_b.
\label{Lagran}
\end{eqnarray}
Here, $F_{\mu\nu}^a=\nabla_\mu A_\nu - \nabla_\nu A_\mu
-ig (T^a)^{bc} A^b_\mu A^c_\nu$, \quad $\bar
D^{ab}_{\mu}=\delta^{ab}\nabla_{\mu}-ig(T^c)^{ab} \bar A^c_{\mu}$,
$T^a$ are the $SU(2)$-group generators taken in the adjoint
representation,  \quad
$A^a_\mu=\bar A^a_\mu +a^a_\mu$, where $\bar A^a_\mu$ is the background field,
$a^a_\mu$ are quantum fluctuations of the gluon field about the background, and
$\chi$ and $\bar \chi$ are ghost fields.

We expand the Lagrangian and keep the terms that are quadratic in gluon
fluctuations, which corresponds to the one-loop approximation. After
this,  the path integral becomes Gaussian and for the effective action
$\Gamma$, related to $Z$ by $Z=\exp (\Gamma)$, we obtain $\Gamma^{(1)}
= \Gamma^{(1)}_4 + \Gamma^{(1)}_2$, where
\begin{eqnarray}
\Gamma^{(1)}_4[\bar A]=
\frac{1}{2} {\rm Tr}\ln [\bar \Theta^{ab}_{\mu\nu}]
-{\rm Tr}\ln[(-\bar D^2)^{ab}], \quad  \mu, \nu \in S^1 \times
R^3,
\label{1}
\end{eqnarray}
(here the first term corresponds to the gluon contribution and the second one is
the ghost contribution), and
\begin{eqnarray}
\Gamma^{(1)}_2[\bar A]=
\frac{1}{2} {\rm Tr}\ln [\bar \Theta^{ab}_{\mu\nu}],\quad  \mu, \nu \in
S^1 \times R^1.
\label{1a}
\end{eqnarray}
The operator $\Theta$ in (\ref{1}) is
defined as
\begin{eqnarray}
\bar \Theta^{ab}_{\mu\nu}=-g_{\mu\nu}(\bar
D_{\lambda} \bar D^{\lambda})^{ab} +2ig(T^c)^{ab} \bar F^c_{\mu\nu}+
(1-1/\xi)(\bar D_{\mu} \bar D_{\nu})^{ab},
\label{2}
\end{eqnarray}
where $T^a$ are the $SU(2)$-group generators taken in the adjoint
representation.  We choose the gauge  $\xi=1$, which makes the third
term in (\ref {2}) vanish. We take the external field potential to be Abelian-
like
\begin{eqnarray}
\bar A^a_{\mu}=n^a \bar A_{\mu},\quad \bar F_{\mu\nu}^a=n^a \bar F_{\mu\nu},
\label{3}
\end{eqnarray}
where $n^a$ is a unit vector pointing in a certain direction in the
color space.

Let $\nu^a\quad (a=1,2,3)$ denote the eigenvalues of the color
matrix $(n^cT^c)$, then we have $|\nu^a|=(1,1,0)$. Finally, for the
operator (\ref{2}) we
obtain:
\begin{eqnarray}
\bar \Theta^{ab}_{\mu\nu}=-g_{\mu\nu}{[\nabla_{\lambda}-ig(n^c T^c)
\bar A_{\lambda}]^2}^{ab}
+ig(n^c T^c)^{ab} \bar F_{\mu\nu}.
\label{op1}
\end{eqnarray}
Following the ideas of \cite{reineng}, in order to model the vortex
configuration in continuum Yang-Mills theory, we take the external
field potential to be pure gauge on the surface of the cylinder
$\rho=const=R$,  i.e.,
$$
\bar A_{\mu} =\frac i g \omega \partial
_{\mu}\omega ^{-1}.
$$
As a simple nonsingular configuration (corresponding to a
"thick vortex" \cite{reineng}),  we  take the gauge field potential
in the $S^1$ circle $\rho=const=R,$ in the polar coordinates, in the
 form
\begin{eqnarray}
A_\mu= \left\{ \begin{array}{cc} \bar
A_{\theta}=\Phi/(2\pi gR),\quad \mu \in S^1\quad(\mu = \theta)\\
0,\quad \mu \in {\bf R}^3\quad (\mu = 2,3,4), \end{array} \right.
\label{ahar}
\end{eqnarray}
with $\Phi /g={\mbox const}$ as
the flux through the cylinder.  As is well known, in this situation we
can introduce the 2-dimensional vector in the $\rho \theta$ plane
$$
G_{\mu}=\frac 1 {2\pi} \varepsilon_{\mu\nu}A_{\nu}.
$$
Then
$$
2\pi r_{\mu}G_{\mu}=\frac i g \omega \frac {\partial}{\partial \theta}
\omega^{-1},
$$
and we have
\begin{eqnarray}
g \ints_0^{2\pi} d\theta
r_{\mu}G_{\mu}=\frac{i}{2\pi}\ints_0^{2\pi}d\theta\omega \frac
{\partial}{\partial \theta}\omega^{-1}=n,
\label{pontr}
\end{eqnarray}
i.e., the topological charge corresponding to mapping from the circle around the
cylinder surface to the $U(1)$ group space (Pontryagin index). In our example
(\ref{ahar}), $\Phi=2\pi n$.

\section{Effective potential}

The spectrum of operator (\ref{op1}) after diagonalization in the
4-dimensional case
\begin{eqnarray}
\bar \Theta^a_{\mu} \equiv -g_{\mu\mu}\Delta^a
\label{common}
\end{eqnarray}
is given by the eigenvalues of the operator
\begin{eqnarray}
\Delta^a \equiv (\nabla_\lambda-ig\nu^a \bar A_\lambda)^2,
\hspace{1cm}
\label{eigop2}
\end{eqnarray}
which are evident in this case:
\begin{eqnarray}
\frac{1}{\rho^2}(l-\nu^a x)^2+{\bf k}^2 \equiv
\Lambda^a_\mu ({\bf k}^2,l), \quad l \in {\bf Z}, \quad \mu= \theta, 2, 3, 4,
\label{eigval}
\end{eqnarray}
where $x = Rg \bar A_{\theta}=\Phi/(2\pi)$.
The same eigenvalue $\Lambda^a
({\bf k}^2,l)= \Lambda^a_\mu ({\bf k}^2,l)$ is obviously obtained for the
corresponding ghost operator in (\ref{1}).

After all the transformation described above we  arrive at the final
formula for the 4-dimensional effective action
\begin{eqnarray}
\Gamma_4^{(1)}[\bar A]=\frac{\Omega}{2\pi R L^3}
\sum_a |\nu^a| \sum_{l=-\infty}^\infty \sum_{{\bf k}}
\left( \frac{1}{2} \sum_{\mu= \theta, 2, 3, 4}
\log[\Lambda_\mu^a({\bf k}^2,l)]- \log[\Lambda^a({\bf k}^2,l)] \right),
\label{effpot}
\end{eqnarray}
 where $\Omega \equiv \int d^4 x=2\pi R \int dx_2
 dx_3 dx_4 $ is the 4-volume of the space time.

In what follows we use the ``proper time'' representation
   \begin{eqnarray}
\log A=-\int\limits_0^\infty \frac{ds}{s}\exp(-sA), {\mbox Re} A>0,
\label{r26}
\end{eqnarray}
which is valid once the subtraction is performed at the lower
integration limit. Next we transform summation over  $l$ in
 (\ref{effpot}) with the help of the identity
\begin{eqnarray}
  \sum_{l=-\infty}^{+\infty}\exp\left[-\frac{s}{R ^2}(l+x)^2\right]
  =\frac{\sqrt {\pi}R}{\sqrt{s}} \sum_{l=1} ^\infty
  \exp(-\frac{\pi^2R^2 l^2}{s}) \cos(2\pi x l).
\label{r28}
\end{eqnarray}
After performing necessary operations in
(\ref{effpot}) we obtain the 4-dimensional
effective potential:
\begin{eqnarray}
  v_4 = -\frac {\Gamma^{(1)}[\bar A]}{\Omega} = -
\frac{1}{2\pi^2}\int\limits_0^\infty \frac{ds}{s^3}\sum_{l=1}^\infty
 \exp(-
\frac{\pi^2R^2 l^2}{s}) \cos(2\pi x l).
\label{r29}
\end{eqnarray}

The final integration over $s$ is easily performed
and we find:
\begin{eqnarray}
  v_{4}= -\frac{4}{\pi^2(2\pi R)^4}\sum_{l=1}^\infty
  \frac{\cos(2\pi x l)}{l^4}=
\frac{4\pi^2}{3(2\pi R)^4} B_4(x).
\label{r30}
\end{eqnarray}
This formula, after replacements $2\pi R$ by $1/T$, and $A_{\theta}$ by
$A_0$, corresponds to the well known finite temperature ($T \ne 0$) result for
the free energy of gluons \cite{weiss} in the background of a constant
$A_0$ potential (see also, \cite{debert}, \cite{engel}).  Similar
calculations for the contribution of the 2-dimensional cylinder give
the result
\begin{eqnarray}
v_{2}= -\frac{2}{\pi
  (2\pi R)^2}\sum_{l=1}^\infty \frac{\cos(2\pi x l)}{l^2}=
-\frac{2\pi}{(2\pi R)^2} B_2(x).
\label{r31}
\end{eqnarray}
Here, and in (\ref{r30}) we used the Bernoulli polinomials
defined according to
$$
\sum_{l=1}^\infty
  \frac{\cos(2\pi lx)}{l^{2n}}=(-1)^{n-1}\frac 1 2
  \frac{(2\pi)^{2n}}{(2n)!}B_{2n}(x).
$$
In particular
$$
B_2(x)=x^2-x+1/6, \nonumber \ B_4=x^4-2x^3+x^2-\frac {1}{30}.
$$
Concerning the results obtained, it should be remarked that,
first, the Bernoulli
polinomials depend on
the argument defined modulo 1, and hence the effective potential conserves the
$Z_2$ symmetry, characteristic for vortices, and, second, evident
renormalization of the result has been performed.
After
summation of the two contributions is performed we obtain
\begin{eqnarray}
v=v_4+\frac{1}{3\pi R^2}v_2.
\label{total}
\end{eqnarray}
The second term for the 2-dimensional cylinder has been averaged over
three possible orientations of the coordinate axes with respect to the cylinder
in
the 4-dimensional space time and over the
area of its cross section. The final result, with the terms independent of
$gA_{\theta}$ having been
omitted, has the simple form
\begin{eqnarray}
v=\frac{1}{12\pi^2R^4}x(2-x)(1-x^2),\quad
x=\frac{\Phi}{2\pi}.
\label{final}
\end{eqnarray}
The result obtained coincides formally with that derived in
\cite{diak} for the following configuration of the 4-dimensional space time:
a plane with a deleted point at the origin, and the
potential $A_{\mu}$ with a singularity at a string going through the
origin. This coincidence has a simple explanation: in both cases we
have the same general topological situation, defined by the Pontryagin index
(\ref{pontr}). It should be mentioned, that in our case the background
potential was defined on the cylinder and had no singularity, while in
the latter case, it was defined on a plane with a singularity at the
origin, and hence the complete basis with Bessel functions had to be
employed in calculating the functional trace.

As in \cite{diak}, the potential
has minima (equal to zero) at $x=0$ and $x=1$, and, due to its periodic
properties, at all
integer values of $x$, though, the potential has a jump of its derivative at
these
points.
The values of the flux that
provide minimum for the effective potential, correspond to $\pm 1$ values
of the elementary Wilson loop that goes along the contour $C$ encircling the
cylinder:
\begin{eqnarray}
 <W(C)>=\frac 1 2 <\Tr~ {\mbox P}~
\exp~ ig\int_C A_{\mu}dx_{\mu}>= \cos (\pi x)
\label{wilson}
\end{eqnarray}
The same situation has already been
reported in \cite{diak}.
As was argued in recent publications (see, e.g., \cite{langf},
\cite{lang}), the value $W(C)= - 1$
characterizes  the vortices, that pierce the area of the large Wilson
loop with Gaussian distribution, thus providing for the area law and for the
linear confining potential.

\section{Acknowledgements}
The author  gratefully acknowledges the hospitality
extended to  him by the theory group of the Institut f\"ur
Theoretische Physik, Universit\"at  T\"ubingen   during his stay
there. This work has been supported in part by the DFG grant 436 RUS 113/477/0.

\end{document}